\documentstyle[prd,aps,epsf]{revtex}

\def\bra#1{\mathopen{\langle#1\,|}}
\def\ket#1{\mathclose{|\,#1\rangle}}

\begin{document}
\author{S. Ying\thanks{
E-Mail: sqying@fudan.edu.cn}}
\author{Department of Physics, Fudan University, Shanghai 200433, China}
\title{Local Finite Density Theory, Statistical Blocking and Color 
Superconductivity\thanks{Talk given at the TMU-Yale Symposium on the Dynamics 
of Gauge Fields, Tokyo, Japan, Dec. 13--15, 1999.} }
\maketitle

\begin{abstract}
The motivation for the development of a local finite density theory is
discussed. One of the problems related to an instability in the baryon
number fluctuation of the chiral symmetry breaking phase of the quark
system in the local theory is shown to exist. Such an instability
problem is removed by taking into account the statistical blocking
effects for the quark propagator, which depends on a macroscopic {\em
statistical blocking parameter } $\varepsilon$. This new frame work is
then applied to study color superconducting phase of the light quark
system.
\end{abstract}

\section{Introduction}

\begin{figure}[h]
\epsfbox{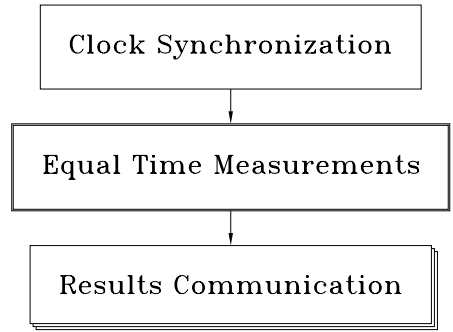}
\caption{The process of local measurement in the relativistic space-time. In
the first and last steps, the observers remain causally (in a classical
sense) connected. The observers are separated by space-like distances in the
second step when the ``equal time'' measurements are carried out. The
correlation between these observers in the second step can only be quantum
mechanical, which is known to violates classical causality manifested in the
violation of Bell's inequality.}
\label{fig:RLM}
\end{figure}

The problems encountered in relativistic high energy physics and
astrophysics are local problems about causally unconnected parts (in a
classical relativistic sense) of the system rather than about the whole
system. In both cases, the resolution of probing system is much smaller than
the size of the systems being observed. The overall feature of the system at
any given time (in any specific reference frame) is generated by first
synchronizing clocks, next measuring wanted physical quantity within each
part of the system {\em independently} at the time that is {\em considered}
the same within the {\em allowed accuracy} and then reassemble the whole
picture of the system using the bits of local information about the system
measured (see Fig. \ref{fig:RLM}). 
There is no guarantee in quantum world that the refined image of the
system obtained under such a procedure should be identical to that of
the one obtained by doing a low resolution global measurement, in
which a uniform (in space-time) external field is exerted on the
system and the response of the system to the external field is
measured, like the ones that determine the bulk properties of the
system frequently studied in non-relativistic condensed matter
system. Such a subtle difference is believed to originate from the
quantum interference effects which are best illustrated by the
classical double slit experiments where when one does not care which
slit the microscopic particle (e.g., an electron) go, there is a
interference pattern on the screen behind the double slits, but as
long as one determines which slit the microscopic particle went
through, the interference pattern disappears. {\em The interference
pattern can not be obtained by summing over the two pictures obtained
separately by knowing exactly which slit the microscopic particle
go}. Such a behavior is no longer a theoretical conjecture but a
experimental fact which is demonstrated recently by experiments from
different groups around the world. The same difference should
manifest in quantum field theory in the local versus global
observables. The importance of an understanding of such a difference
has not been appreciated due to the lack of a comparing theoretical
frame-work. A consistent local theory for quantum field theory at
finite field strength (leading to finite density of matter and energy)
is constructed \cite {fd-th1,fd-th2} not only for understand such a
difference but also for describing the problems raised in high energy
physics and astrophysics.

I shall discuss one of the problems in constructing a local finite
density quantum field theory for fermions in this talk, those who are
interested in the original works can read the papers
\cite{fd-th1,fd-th2}. The particular problem that I am going to to
discuss here is about the necessity of including fermionic statistical
block effects in the chiral symmetry breaking phase of light quark
system. But before we can discuss the problem, let us first brief
review what the local theory should be look like at the formal level
and what is its most transparent properties.

\section{The Statistical Gauge Field}

The non-relativistic many body theory for a system with variable particle
number is based upon the Grand Canonical Ensemble (GCE) \cite{Huang}. The
partition function for the system in the GCE is related to its Hamiltonian $
\widehat H$ and its particle number $\widehat N$ in the following way 
\begin{equation}
Z = \lim_{\Omega\to\infty } Tr e^{-\beta(\widehat H - \mu \widehat N)},
\label{part-func}
\end{equation}
where $\Omega$ is the volume of the system, $\beta = 1/T$ with $T$ the
temperature and $\mu$ is the chemical potential. The limit $\Omega\to\infty$
is the thermodynamical limit in which the equilibrium statistical mechanics
describes physical many body system at equilibrium. A measurement of the
ground state fermion number density in GCE which is given by 
\begin{eqnarray}
\overline n &=& \lim_{T\to 0}{\frac{\partial}{\partial \mu}} \left ({\frac{T
}{\Omega}} \ln Z \right )  \label{gl-rho}
\end{eqnarray}
corresponds to a global measurement since the chemical potential is
independent of both space and time. Global measurements are most
frequently carried out in condensed matter systems with a size smaller
than or comparable to our human body (of order 1 meter) which permit
us to study the system ``as a whole''. The GCE is highly successful in
describing these situations. Questions posed by high-energy physics,
astrophysics and cosmology are somewhat different. The size of the
systems compared to that of the resolution of the measuring apparatus
are normally much larger, therefore one actually is capable of study
either more details of the probed systems due to the reduction of the
wave length of the probing system or only small fraction of the probed
system due to the impossibility of covering the whole probed system by
the probing system at the {\em same time}. One is in effect doing
local measurements in these later situations.

Local measurements in quantum field theory are realized by exerting an
external local field to the system at the space-time points interested, and,
the measured quantity is deduced from the response of the system to the
external field. For finite density systems, a Lorentz 4-vector local field $
\mu^\alpha(x)$, called the primary statistical gauge field, is introduced.
The ground state expectation value of fermion number density can be
expressed as the functional derivative 
\begin{eqnarray}
\overline \rho(x) &=& {\frac{\delta \ln Z}{\delta\mu^0(x)}}  \label{lc-rho}
\end{eqnarray}
with $Z$ a functional of $\mu^\alpha$. It corresponds to a local measurement
in the ground state at the space-time point x.

For a concrete discussion of the consequences of substituting
the global chemical potential in non-relativistic many body theory for the
local primary statistical gauge field, let us consider an Lagrangian of the
following form 
\begin{eqnarray}
{\cal L}^{\prime}&=& {\frac{1}{2}}\overline\Psi (i\rlap\slash\partial + 
\rlap\slash\hspace{-1pt}\mu O_3 - m) \Psi + {\cal L}_B + {\cal L}_{
\mbox{\scriptsize int}}
\end{eqnarray}
with $m$ the fermion mass, ${\cal L}_B$ the Lagrangian density of boson
fields of the system and ${\cal L}_{\mbox{\scriptsize int}}$ the interaction
between the boson and fermion fields. An 8 component ``real'' fermion field $
\Psi$ is used for the discussion \cite{fd-th1} with $O_3$ the third Pauli
matrices acting on the upper and lower 4-component of $\Psi$. In most of the
cases, ${\cal L}_{\mbox{\scriptsize int}}=-\overline\Psi(\Sigma(f)-m)\Psi$
with $\Sigma$ a function(al) of the boson fields represented by $f$. It is
well known that the partition functional for the system can be formally
written in terms of a path integration over the dynamical fields 
\begin{equation}
Z[\mu]=\int D[f,\Psi,\overline\Psi] exp(i\int d^4 x{\cal L}^{\prime})= \int
D[f] exp(iS_{\mbox{\scriptsize eff}})
\end{equation}
which is a functional of $\mu^\alpha$ and $\ln Z[\mu]$ is 
its effective action.

The fermion degrees of freedom can be integrated out first leading to an
effective Euclidean action for the boson fields and the primary statistical
gauge field 
\begin{eqnarray}
S_{\mbox{\scriptsize eff}}[f,\mu] &=& {\frac{1}{2}} \mbox{SpLn} {\frac{ [i
\rlap\slash\partial + \rlap\slash\hspace{-1pt}\mu O_3-\Sigma(f)]}{[i
\rlap\slash\partial + \rlap\slash\hspace{-1pt}\mu O_3-m]}} + \int d^4 x 
\left [{\cal L }_B(f) + \mu\overline\rho-\overline e \right ]
\label{eff-action}
\end{eqnarray}
with ``Sp'' the functional trace, $\overline e$ the average energy density
of the corresponding free system of fermions ($\Sigma=0$) and $\overline\rho$
is the ground state expectation value of fermion number density to be
discussed in the following at the ``chemical potential'' $\mu\equiv\sqrt{
\overline\mu^\alpha \overline\mu_\alpha}$, where $\overline \mu^\alpha$ is
the ground state value of $\mu^\alpha$ normally provided by the external
conditions that fixes the density.

The first term in Eq. \ref{eff-action} is invariant under the $U(1)$
statistical gauge transformation $\mu_\alpha(x)\to \mu^{\prime}_\alpha(x)=
\mu_\alpha(x)-\partial_\alpha\Lambda(x)$. This is because such a gauge
transformation, which transforms the partition functional $Z[\mu]$ to 
\begin{equation}
Z[\mu] \to Z^{\prime}[\mu] = Z[\mu^{\prime}]
\end{equation}
can always be compensated, at least for the first term in Eq. \ref
{eff-action} by a change of the fermionic integration variable 
\begin{eqnarray}
\Psi(x) &\to & e^{i\Lambda(x) O_3} \Psi(x)  \label{G-tran1} \\
\overline \Psi(x) &\to & e^{-i\Lambda(x) O_3} \overline \Psi(x)
\label{G-tran2}
\end{eqnarray}
so that 
\begin{equation}
Z[\mu^{\prime}] = Z[\mu],
\end{equation}
which means that fermionic quantum fluctuation contribution to $Z[\mu]$ is
invariant under the gauge transformation. Such a gauge invariance is
connected to the conservation of fermion number. Therefore I call
the 4-vector $\mu^\alpha$ the statistical gauge field. Since
$\mu^\alpha$ is a local field, its excitation represents certain
collective excitations of the system. Its general dynamics in
different phases of the system at low energy can be derived
\cite{fd-th1}.

\section{The Instability of GCE in the Relativistic Space-Time}

The Euclidean effective action for the boson field given by Eq. \ref
{eff-action} can be evaluated in the usual way \cite{fd-th1}. Since the
effective action given by Eq. \ref{eff-action} is a canonical functional of $
\mu^\alpha$, we can make a Legendre transformation of it, namely 
\begin{equation}
\widetilde S_{\mbox{\scriptsize eff}} = S_{\mbox{\scriptsize eff}} -\int d^4
x \overline \mu_\alpha \overline j^\alpha
\end{equation}
with $\overline j^\alpha$ the ground state current of fermion number
density, to make it a canonical functional of the fermion density in order
to study the stability of the vacuum state against fluctuations in
fermion number.

The effective potential characterizing the above mentioned energy density
can be defined using $\widetilde S_{\mbox{\scriptsize eff}}$ for space-time
independent background $f$ fields: 
\begin{equation}
V_{\mbox{\scriptsize eff}}=-\widetilde S_{\mbox{\scriptsize
eff}}/V_4
\end{equation}
with $V_4$ the volume of the space-time box that contains the system. The
stability of the vacuum state against the fluctuations in fermion number
density around $\overline\rho=0 $ can be studied using $V_{\mbox{\scriptsize
eff}}$, which is a canonical function of $\overline\rho$. Minimization of $
V_{\mbox{\scriptsize
eff}} $ with respect to $\mu$ gives the vacuum value $\rho_{vac}$ of the
system since the vacuum $\overline\rho$ is a monotonic function of $\mu$.

The local finite density theory can be examined by applying it to the chiral
symmetry breaking phase of the half bosonized Nambu Jona-Lasinio model for
the 3+1 dimension and the chiral Gross-Neveu model for 2+1 dimensions with
its effective potential given by 
\begin{equation}
V_{\mbox{\scriptsize eff}} = i N_g \int {\frac{d^Dp}{(2\pi)^D}}\left [ \ln
\left ( 1-{\frac{\sigma^2 }{p_+^2}} \right ) + \ln \left ( 1-{\frac{\sigma^2 
}{p_-^2}} \right ) \right ] + {\frac{1}{4 G_0}} \sigma^2 + \overline e,
\label{Veff1}
\end{equation}
where ``$D$'' is the space-time dimension, $G_0$ is the coupling constant
and $p^2_\pm = (p_0\pm \mu)^2-\mbox{\boldmath{p}}^2$.

The integration over the momentum $p^\mu$ in Eq. \ref{Veff1} is divergent, a
cutoff is needed to regulate the theory so that finite results can be
obtained. Whatever the cutoff scheme is, one qualitative feature is true,
namely, the stable minima of $V_{eff}$ move from $\sigma=0$ to $\sigma = \pm
\sigma_{vac}\ne 0$ as the coupling constant $G_0$ is increased above a
critical value $G_{0c}$. 

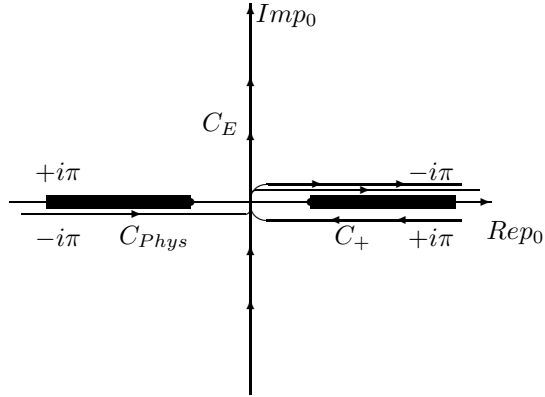
\begin{figure}[h]
\unitlength=0.80mm
\linethickness{0.5pt}
\begin{picture}(120.00,70.00)(-27,42)
\put(40.00,80.00){\vector(1,0){80.00}}
\put(124.00,75.00){\makebox(0,0)[cc]{$Rep_0$}}
\put(86.00,111.00){\makebox(0,0)[cc]{$Imp_0$}}
\put(90.00,78.95){\rule{24.00\unitlength}{2.10\unitlength}}
\put(46.00,78.95){\rule{24.00\unitlength}{2.10\unitlength}}
\put(90.00,80.00){\circle*{1.50}}
\put(70.00,80.00){\circle*{1.50}}
\put(42.00,78.00){\line(1,0){36.00}}
\put(78.00,79.00){\oval(4.00,2.00)[rb]}
\put(82.50,81.00){\oval(5.00,2.00)[lt]}
\put(82.00,82.00){\line(1,0){36.00}}
\put(96.00,82.00){\vector(1,0){4.00}}
\put(57.00,78.00){\vector(1,0){5.00}}
\put(80.00,48.00){\vector(0,1){65.00}}
\put(80.00,62.00){\vector(0,1){2.00}}
\put(80.00,71.00){\vector(0,1){2.00}}
\put(80.00,90.00){\vector(0,1){2.00}}
\put(80.00,98.00){\vector(0,1){3.00}}
\put(83.00,83.00){\line(1,0){32.00}}
\put(83.00,77.00){\line(1,0){32.00}}
\put(83.50,80.00){\oval(7.00,6.00)[l]}
\put(90.00,83.00){\vector(1,0){2.00}}
\put(104.00,83.00){\vector(1,0){2.00}}
\put(106.00,77.00){\vector(-1,0){2.00}}
\put(95.00,77.00){\vector(-1,0){2.00}}
\put(64.00,74.00){\makebox(0,0)[cc]{$C_{Phys}$}}
\put(75.00,93.00){\makebox(0,0)[cc]{$C_E$}}
\put(97.00,74.00){\makebox(0,0)[cc]{$C_+$}}
\put(110.00,85.00){\makebox(0,0)[cc]{$-i\pi$}}
\put(110.00,74.00){\makebox(0,0)[cc]{$+i\pi$}}
\put(48.00,74.00){\makebox(0,0)[cc]{$-i\pi$}}
\put(48.00,85.00){\makebox(0,0)[cc]{$+i\pi$}}
\end{picture}
\caption{The set of contours belonging to the same physical class. Contour $
C_{Phys}$ is the original physical contour in the Minkowski space. Contour $
C_E$ is the Euclidean contour. Contour $C_+$ is the quasiparticle contour. $
\pm i\pi$ denote the imaginary part of the integrand along the edges of its
cuts (thick lines) on the physical $p_0$ plane.}
\label{Fig:Cont1}
\end{figure}

Since the logarithmic functions in Eq. \ref{Veff1}
has branch cuts on the real $p^0$ axis, the choice for the integration
contour for $p^0$ is important. The physical $p^0$ integration contour is
shown by $C_{Phys}$ in Fig. \ref{Fig:Cont1}, which goes under the real axis
when $p^0< 0$ and above the real axis when $p^0>0$. 

It is possible to do the momentum space integration by {\em doing the $p^0$
integration first} without a cutoff in it, namely, 
\begin{equation}
V_{\mbox{\scriptsize eff}}=iN_g\int {\frac{d^{D-1}p}{(2\pi )^{D-1}}}
\int_{C_{Phys}}{\frac{dp^0}{2\pi }}\left[ \ln \left( 1-{\frac{\sigma ^2}{
p_{+}^2}}\right) +\ln \left( 1-{\frac{\sigma ^2}{p_{-}^2}}\right) \right] +{
\frac 1{4G_0}}\sigma ^2+\overline{e}.\label{Veff2}
\end{equation}
In such a case, the integration contours $C_{Phys}$, $C_{+}$ and $C_E$
are all equivalent in the 8-component ``real'' representation for the
fermion fields, since the large $p_0$ behavior of the imaginary part
of the logarithmic functions in Eq. \ref{Veff2} is of order $O(\mu
^2/|p_0|^2)$ on the physical complex $p_0$ sheet, which allows the
distortion of the integration contour from $C_{Phys}$ to the Euclidean
contour $C_E$ and to the quasiparticle contour $C_{+}$. Let us make a
few comment: 1) the 8-component ``real'' representation for the
fermion fields make it equivalent between contour $C_{Phys}$, $C_{+}$
and $C_E$. The usual 4-component representation does not has such a
property \cite{pathint} without an arbitrary subtraction. It is in
this sense that the 8-component theory is not equivalent to the
4-component theory if finite density situation is considered 2) the
quasiparticle contour $C_{+}$ is the most convenient contour to
demonstrate that the quantum fluctuation term (the fermionic
determinant) at the right hand side of Eq. \ref{Veff2} is a sum over
the quasiparticle energies levels in the negative energy sea.  The
behavior of the quasi-particles is, at least in the approximation
taken here, identical to that of genuine particles with a global mass
despite the fact that the ``mass'' of the quasi-particles is related
to the vacuum expectation value of a local field rather than a global
constant 3) there is no instability due the fermion number fluctuation
if the $p^0$ integration is done first since only the quasiparticle
contributes. This integration contour for $p^0$ is implied in most of
the phenomenological applications in the literature since it is
intuitively acceptable. Such a ``naturalness'' is based on an implicit
assumption, namely, the quasiparticles is the whole story.

However, there are a few problems in doing a unbounded $p^0$
integration first in a relativistic space-time. The most obvious one
is related to the fact that a unbounded $p^0$ integration, which is
done first, corresponds to an arbitrary choice of a equal-time
hypersurface in space-time with zero thickness. This is not a
covariant choice since the spatial component of the momentum $p^\mu$
integration has to be cutted off so that the spatial resolution of the
theory is not infinitely high. Such a procedure, which is Newtonian in
character based on a universal absolute time, is not considered fair
in relativity since space and time can be related to each other by
Lorentz transformation. The second one is connected to the observation
that the quasiparticle does not saturate the local observables
\cite{fd-th1,quasip} due to the existence of the transient and local
quantum fluctuations of the order parameter $\sigma$. These important
kind of transient and local quantum fluctuations, which does not
decouple in the thermodynamical limit\cite{quasip}, are quenched if
the momentum $ p^\mu$ integration is carried out by doing an unbounded
$p^0$ integration first.

The relativistic covariance can be maintained by cutting off the theory 
covariantly in the Euclidean space for the momentum $p^\mu$ by restricting
the length of the Euclidean $p^\mu$ to be less than or equal to a cutoff
$\Lambda\sim 1$ $GeV$. Such a covariance in cutoff is also adopted in lattice
simulations of QCD on a symmetric lattice. The covariant cutoff of the theory
allow the observers on each space-time point of the measuring hypersurface
to have a resolution in time (speed of light $c=1$) of the same order as
the spatial resolution. Therefore we are not talking about absolute
equal-time, which does not make sense in relativity, but the kind of
``equal-time'' defined operationally in Fig. \ref{fig:RLM}.
It is not expected that the
difference between these two kind of cutoff procedure can be removed
by renormalization since such differences are physical that shall not
be renormalized away.

\begin{figure}[h]
\begin{center}
\setlength{\unitlength}{0.240900pt}
\ifx\plotpoint\undefined\newsavebox{\plotpoint}\fi
\sbox{\plotpoint}{\rule[-0.500pt]{1.000pt}{1.000pt}}%
\begin{picture}(1500,900)(0,0)
\font\gnuplot=cmr10 at 10pt
\gnuplot
\sbox{\plotpoint}{\rule[-0.500pt]{1.000pt}{1.000pt}}%
\put(220.0,113.0){\rule[-0.500pt]{4.818pt}{1.000pt}}
\put(198,113){\makebox(0,0)[r]{$-1.4$}}
\put(1416.0,113.0){\rule[-0.500pt]{4.818pt}{1.000pt}}
\put(220.0,266.0){\rule[-0.500pt]{4.818pt}{1.000pt}}
\put(198,266){\makebox(0,0)[r]{$-1.36$}}
\put(1416.0,266.0){\rule[-0.500pt]{4.818pt}{1.000pt}}
\put(220.0,419.0){\rule[-0.500pt]{4.818pt}{1.000pt}}
\put(198,419){\makebox(0,0)[r]{$-1.32$}}
\put(1416.0,419.0){\rule[-0.500pt]{4.818pt}{1.000pt}}
\put(220.0,571.0){\rule[-0.500pt]{4.818pt}{1.000pt}}
\put(198,571){\makebox(0,0)[r]{$-1.28$}}
\put(1416.0,571.0){\rule[-0.500pt]{4.818pt}{1.000pt}}
\put(220.0,724.0){\rule[-0.500pt]{4.818pt}{1.000pt}}
\put(198,724){\makebox(0,0)[r]{$-1.24$}}
\put(1416.0,724.0){\rule[-0.500pt]{4.818pt}{1.000pt}}
\put(220.0,877.0){\rule[-0.500pt]{4.818pt}{1.000pt}}
\put(198,877){\makebox(0,0)[r]{$-1.2$}}
\put(1416.0,877.0){\rule[-0.500pt]{4.818pt}{1.000pt}}
\put(220.0,113.0){\rule[-0.500pt]{1.000pt}{4.818pt}}
\put(220,68){\makebox(0,0){$-0.4$}}
\put(220.0,857.0){\rule[-0.500pt]{1.000pt}{4.818pt}}
\put(524.0,113.0){\rule[-0.500pt]{1.000pt}{4.818pt}}
\put(524,68){\makebox(0,0){$-0.2$}}
\put(524.0,857.0){\rule[-0.500pt]{1.000pt}{4.818pt}}
\put(828.0,113.0){\rule[-0.500pt]{1.000pt}{4.818pt}}
\put(828,68){\makebox(0,0){$0$}}
\put(828.0,857.0){\rule[-0.500pt]{1.000pt}{4.818pt}}
\put(1132.0,113.0){\rule[-0.500pt]{1.000pt}{4.818pt}}
\put(1132,68){\makebox(0,0){$0.2$}}
\put(1132.0,857.0){\rule[-0.500pt]{1.000pt}{4.818pt}}
\put(1436.0,113.0){\rule[-0.500pt]{1.000pt}{4.818pt}}
\put(1436,68){\makebox(0,0){$0.4$}}
\put(1436.0,857.0){\rule[-0.500pt]{1.000pt}{4.818pt}}
\put(220.0,113.0){\rule[-0.500pt]{292.934pt}{1.000pt}}
\put(1436.0,113.0){\rule[-0.500pt]{1.000pt}{184.048pt}}
\put(220.0,877.0){\rule[-0.500pt]{292.934pt}{1.000pt}}
\put(45,495){\makebox(0,0){$V_{eff}/\Lambda^4\hspace{0.3cm}$}}
\put(828,23){\makebox(0,0){$\mu/\Lambda$}}
\put(828,571){\makebox(0,0){$\alpha_0 = 0.8$}}
\put(281,915){\makebox(0,0){$\times 10^{-2}$}}
\put(220.0,113.0){\rule[-0.500pt]{1.000pt}{184.048pt}}
\sbox{\plotpoint}{\rule[-0.175pt]{0.350pt}{0.350pt}}%
\put(294,728){\usebox{\plotpoint}}
\multiput(294.48,720.48)(0.502,-2.562){25}{\rule{0.121pt}{1.812pt}}
\multiput(293.27,724.24)(14.000,-65.238){2}{\rule{0.350pt}{0.906pt}}
\multiput(308.48,651.60)(0.502,-2.527){23}{\rule{0.121pt}{1.784pt}}
\multiput(307.27,655.30)(13.000,-59.298){2}{\rule{0.350pt}{0.892pt}}
\multiput(321.48,589.49)(0.502,-2.204){23}{\rule{0.121pt}{1.568pt}}
\multiput(320.27,592.74)(13.000,-51.745){2}{\rule{0.350pt}{0.784pt}}
\multiput(334.48,535.45)(0.502,-1.852){25}{\rule{0.121pt}{1.337pt}}
\multiput(333.27,538.22)(14.000,-47.224){2}{\rule{0.350pt}{0.669pt}}
\multiput(348.48,485.83)(0.502,-1.720){23}{\rule{0.121pt}{1.245pt}}
\multiput(347.27,488.42)(13.000,-40.416){2}{\rule{0.350pt}{0.623pt}}
\multiput(361.48,443.69)(0.502,-1.404){25}{\rule{0.121pt}{1.038pt}}
\multiput(360.27,445.85)(14.000,-35.847){2}{\rule{0.350pt}{0.519pt}}
\multiput(375.48,405.95)(0.502,-1.316){23}{\rule{0.121pt}{0.976pt}}
\multiput(374.27,407.97)(13.000,-30.974){2}{\rule{0.350pt}{0.488pt}}
\multiput(388.48,373.62)(0.502,-1.074){23}{\rule{0.121pt}{0.814pt}}
\multiput(387.27,375.31)(13.000,-25.310){2}{\rule{0.350pt}{0.407pt}}
\multiput(401.48,347.25)(0.502,-0.844){25}{\rule{0.121pt}{0.662pt}}
\multiput(400.27,348.62)(14.000,-21.625){2}{\rule{0.350pt}{0.331pt}}
\multiput(415.48,324.51)(0.502,-0.751){23}{\rule{0.121pt}{0.599pt}}
\multiput(414.27,325.76)(13.000,-17.757){2}{\rule{0.350pt}{0.300pt}}
\multiput(428.48,305.96)(0.502,-0.589){23}{\rule{0.121pt}{0.491pt}}
\multiput(427.27,306.98)(13.000,-13.980){2}{\rule{0.350pt}{0.246pt}}
\multiput(441.00,292.02)(0.597,-0.502){21}{\rule{0.496pt}{0.121pt}}
\multiput(441.00,292.27)(12.971,-12.000){2}{\rule{0.248pt}{0.350pt}}
\multiput(455.00,280.02)(0.753,-0.503){15}{\rule{0.593pt}{0.121pt}}
\multiput(455.00,280.27)(11.769,-9.000){2}{\rule{0.297pt}{0.350pt}}
\multiput(468.00,271.02)(1.477,-0.507){7}{\rule{0.998pt}{0.122pt}}
\multiput(468.00,271.27)(10.930,-5.000){2}{\rule{0.499pt}{0.350pt}}
\multiput(481.00,266.02)(3.428,-0.516){3}{\rule{1.721pt}{0.124pt}}
\multiput(481.00,266.27)(10.428,-3.000){2}{\rule{0.860pt}{0.350pt}}
\put(495,262.77){\rule{3.132pt}{0.350pt}}
\multiput(495.00,263.27)(6.500,-1.000){2}{\rule{1.566pt}{0.350pt}}
\put(508,263.27){\rule{2.363pt}{0.350pt}}
\multiput(508.00,262.27)(8.097,2.000){2}{\rule{1.181pt}{0.350pt}}
\multiput(521.00,265.47)(3.428,0.516){3}{\rule{1.721pt}{0.124pt}}
\multiput(521.00,264.27)(10.428,3.000){2}{\rule{0.860pt}{0.350pt}}
\multiput(535.00,268.47)(1.477,0.507){7}{\rule{0.998pt}{0.122pt}}
\multiput(535.00,267.27)(10.930,5.000){2}{\rule{0.499pt}{0.350pt}}
\multiput(548.00,273.47)(1.186,0.505){9}{\rule{0.846pt}{0.122pt}}
\multiput(548.00,272.27)(11.244,6.000){2}{\rule{0.423pt}{0.350pt}}
\multiput(561.00,279.47)(1.073,0.504){11}{\rule{0.787pt}{0.121pt}}
\multiput(561.00,278.27)(12.366,7.000){2}{\rule{0.394pt}{0.350pt}}
\multiput(575.00,286.47)(0.856,0.504){13}{\rule{0.656pt}{0.121pt}}
\multiput(575.00,285.27)(11.638,8.000){2}{\rule{0.328pt}{0.350pt}}
\multiput(588.00,294.47)(0.753,0.503){15}{\rule{0.593pt}{0.121pt}}
\multiput(588.00,293.27)(11.769,9.000){2}{\rule{0.297pt}{0.350pt}}
\multiput(601.00,303.47)(0.813,0.503){15}{\rule{0.632pt}{0.121pt}}
\multiput(601.00,302.27)(12.688,9.000){2}{\rule{0.316pt}{0.350pt}}
\multiput(615.00,312.48)(0.672,0.503){17}{\rule{0.542pt}{0.121pt}}
\multiput(615.00,311.27)(11.874,10.000){2}{\rule{0.271pt}{0.350pt}}
\multiput(628.00,322.48)(0.672,0.503){17}{\rule{0.542pt}{0.121pt}}
\multiput(628.00,321.27)(11.874,10.000){2}{\rule{0.271pt}{0.350pt}}
\multiput(641.00,332.48)(0.725,0.503){17}{\rule{0.577pt}{0.121pt}}
\multiput(641.00,331.27)(12.801,10.000){2}{\rule{0.289pt}{0.350pt}}
\multiput(655.00,342.47)(0.753,0.503){15}{\rule{0.593pt}{0.121pt}}
\multiput(655.00,341.27)(11.769,9.000){2}{\rule{0.297pt}{0.350pt}}
\multiput(668.00,351.48)(0.672,0.503){17}{\rule{0.542pt}{0.121pt}}
\multiput(668.00,350.27)(11.874,10.000){2}{\rule{0.271pt}{0.350pt}}
\multiput(681.00,361.47)(0.813,0.503){15}{\rule{0.632pt}{0.121pt}}
\multiput(681.00,360.27)(12.688,9.000){2}{\rule{0.316pt}{0.350pt}}
\multiput(695.00,370.47)(0.856,0.504){13}{\rule{0.656pt}{0.121pt}}
\multiput(695.00,369.27)(11.638,8.000){2}{\rule{0.328pt}{0.350pt}}
\multiput(708.00,378.47)(0.856,0.504){13}{\rule{0.656pt}{0.121pt}}
\multiput(708.00,377.27)(11.638,8.000){2}{\rule{0.328pt}{0.350pt}}
\multiput(721.00,386.47)(1.073,0.504){11}{\rule{0.787pt}{0.121pt}}
\multiput(721.00,385.27)(12.366,7.000){2}{\rule{0.394pt}{0.350pt}}
\multiput(735.00,393.47)(0.994,0.504){11}{\rule{0.738pt}{0.121pt}}
\multiput(735.00,392.27)(11.469,7.000){2}{\rule{0.369pt}{0.350pt}}
\multiput(748.00,400.47)(1.477,0.507){7}{\rule{0.998pt}{0.122pt}}
\multiput(748.00,399.27)(10.930,5.000){2}{\rule{0.499pt}{0.350pt}}
\multiput(761.00,405.47)(1.595,0.507){7}{\rule{1.067pt}{0.122pt}}
\multiput(761.00,404.27)(11.784,5.000){2}{\rule{0.534pt}{0.350pt}}
\multiput(775.00,410.47)(1.979,0.509){5}{\rule{1.225pt}{0.123pt}}
\multiput(775.00,409.27)(10.457,4.000){2}{\rule{0.613pt}{0.350pt}}
\put(788,414.27){\rule{2.363pt}{0.350pt}}
\multiput(788.00,413.27)(8.097,2.000){2}{\rule{1.181pt}{0.350pt}}
\put(801,416.27){\rule{2.538pt}{0.350pt}}
\multiput(801.00,415.27)(8.733,2.000){2}{\rule{1.269pt}{0.350pt}}
\put(841,416.27){\rule{2.538pt}{0.350pt}}
\multiput(841.00,417.27)(8.733,-2.000){2}{\rule{1.269pt}{0.350pt}}
\put(855,414.27){\rule{2.363pt}{0.350pt}}
\multiput(855.00,415.27)(8.097,-2.000){2}{\rule{1.181pt}{0.350pt}}
\multiput(868.00,413.02)(1.979,-0.509){5}{\rule{1.225pt}{0.123pt}}
\multiput(868.00,413.27)(10.457,-4.000){2}{\rule{0.613pt}{0.350pt}}
\multiput(881.00,409.02)(1.595,-0.507){7}{\rule{1.067pt}{0.122pt}}
\multiput(881.00,409.27)(11.784,-5.000){2}{\rule{0.534pt}{0.350pt}}
\multiput(895.00,404.02)(1.477,-0.507){7}{\rule{0.998pt}{0.122pt}}
\multiput(895.00,404.27)(10.930,-5.000){2}{\rule{0.499pt}{0.350pt}}
\multiput(908.00,399.02)(0.994,-0.504){11}{\rule{0.738pt}{0.121pt}}
\multiput(908.00,399.27)(11.469,-7.000){2}{\rule{0.369pt}{0.350pt}}
\multiput(921.00,392.02)(1.073,-0.504){11}{\rule{0.787pt}{0.121pt}}
\multiput(921.00,392.27)(12.366,-7.000){2}{\rule{0.394pt}{0.350pt}}
\multiput(935.00,385.02)(0.856,-0.504){13}{\rule{0.656pt}{0.121pt}}
\multiput(935.00,385.27)(11.638,-8.000){2}{\rule{0.328pt}{0.350pt}}
\multiput(948.00,377.02)(0.856,-0.504){13}{\rule{0.656pt}{0.121pt}}
\multiput(948.00,377.27)(11.638,-8.000){2}{\rule{0.328pt}{0.350pt}}
\multiput(961.00,369.02)(0.813,-0.503){15}{\rule{0.632pt}{0.121pt}}
\multiput(961.00,369.27)(12.688,-9.000){2}{\rule{0.316pt}{0.350pt}}
\multiput(975.00,360.02)(0.672,-0.503){17}{\rule{0.542pt}{0.121pt}}
\multiput(975.00,360.27)(11.874,-10.000){2}{\rule{0.271pt}{0.350pt}}
\multiput(988.00,350.02)(0.753,-0.503){15}{\rule{0.593pt}{0.121pt}}
\multiput(988.00,350.27)(11.769,-9.000){2}{\rule{0.297pt}{0.350pt}}
\multiput(1001.00,341.02)(0.725,-0.503){17}{\rule{0.577pt}{0.121pt}}
\multiput(1001.00,341.27)(12.801,-10.000){2}{\rule{0.289pt}{0.350pt}}
\multiput(1015.00,331.02)(0.672,-0.503){17}{\rule{0.542pt}{0.121pt}}
\multiput(1015.00,331.27)(11.874,-10.000){2}{\rule{0.271pt}{0.350pt}}
\multiput(1028.00,321.02)(0.672,-0.503){17}{\rule{0.542pt}{0.121pt}}
\multiput(1028.00,321.27)(11.874,-10.000){2}{\rule{0.271pt}{0.350pt}}
\multiput(1041.00,311.02)(0.813,-0.503){15}{\rule{0.632pt}{0.121pt}}
\multiput(1041.00,311.27)(12.688,-9.000){2}{\rule{0.316pt}{0.350pt}}
\multiput(1055.00,302.02)(0.753,-0.503){15}{\rule{0.593pt}{0.121pt}}
\multiput(1055.00,302.27)(11.769,-9.000){2}{\rule{0.297pt}{0.350pt}}
\multiput(1068.00,293.02)(0.856,-0.504){13}{\rule{0.656pt}{0.121pt}}
\multiput(1068.00,293.27)(11.638,-8.000){2}{\rule{0.328pt}{0.350pt}}
\multiput(1081.00,285.02)(1.073,-0.504){11}{\rule{0.787pt}{0.121pt}}
\multiput(1081.00,285.27)(12.366,-7.000){2}{\rule{0.394pt}{0.350pt}}
\multiput(1095.00,278.02)(1.186,-0.505){9}{\rule{0.846pt}{0.122pt}}
\multiput(1095.00,278.27)(11.244,-6.000){2}{\rule{0.423pt}{0.350pt}}
\multiput(1108.00,272.02)(1.477,-0.507){7}{\rule{0.998pt}{0.122pt}}
\multiput(1108.00,272.27)(10.930,-5.000){2}{\rule{0.499pt}{0.350pt}}
\multiput(1121.00,267.02)(3.428,-0.516){3}{\rule{1.721pt}{0.124pt}}
\multiput(1121.00,267.27)(10.428,-3.000){2}{\rule{0.860pt}{0.350pt}}
\put(1135,263.27){\rule{2.363pt}{0.350pt}}
\multiput(1135.00,264.27)(8.097,-2.000){2}{\rule{1.181pt}{0.350pt}}
\put(1148,262.77){\rule{3.132pt}{0.350pt}}
\multiput(1148.00,262.27)(6.500,1.000){2}{\rule{1.566pt}{0.350pt}}
\multiput(1161.00,264.47)(3.428,0.516){3}{\rule{1.721pt}{0.124pt}}
\multiput(1161.00,263.27)(10.428,3.000){2}{\rule{0.860pt}{0.350pt}}
\multiput(1175.00,267.47)(1.477,0.507){7}{\rule{0.998pt}{0.122pt}}
\multiput(1175.00,266.27)(10.930,5.000){2}{\rule{0.499pt}{0.350pt}}
\multiput(1188.00,272.47)(0.753,0.503){15}{\rule{0.593pt}{0.121pt}}
\multiput(1188.00,271.27)(11.769,9.000){2}{\rule{0.297pt}{0.350pt}}
\multiput(1201.00,281.48)(0.597,0.502){21}{\rule{0.496pt}{0.121pt}}
\multiput(1201.00,280.27)(12.971,12.000){2}{\rule{0.248pt}{0.350pt}}
\multiput(1215.48,293.00)(0.502,0.589){23}{\rule{0.121pt}{0.491pt}}
\multiput(1214.27,293.00)(13.000,13.980){2}{\rule{0.350pt}{0.246pt}}
\multiput(1228.48,308.00)(0.502,0.751){23}{\rule{0.121pt}{0.599pt}}
\multiput(1227.27,308.00)(13.000,17.757){2}{\rule{0.350pt}{0.300pt}}
\multiput(1241.48,327.00)(0.502,0.844){25}{\rule{0.121pt}{0.662pt}}
\multiput(1240.27,327.00)(14.000,21.625){2}{\rule{0.350pt}{0.331pt}}
\multiput(1255.48,350.00)(0.502,1.074){23}{\rule{0.121pt}{0.814pt}}
\multiput(1254.27,350.00)(13.000,25.310){2}{\rule{0.350pt}{0.407pt}}
\multiput(1268.48,377.00)(0.502,1.316){23}{\rule{0.121pt}{0.976pt}}
\multiput(1267.27,377.00)(13.000,30.974){2}{\rule{0.350pt}{0.488pt}}
\multiput(1281.48,410.00)(0.502,1.404){25}{\rule{0.121pt}{1.038pt}}
\multiput(1280.27,410.00)(14.000,35.847){2}{\rule{0.350pt}{0.519pt}}
\multiput(1295.48,448.00)(0.502,1.720){23}{\rule{0.121pt}{1.245pt}}
\multiput(1294.27,448.00)(13.000,40.416){2}{\rule{0.350pt}{0.623pt}}
\multiput(1308.48,491.00)(0.502,1.852){25}{\rule{0.121pt}{1.337pt}}
\multiput(1307.27,491.00)(14.000,47.224){2}{\rule{0.350pt}{0.669pt}}
\multiput(1322.48,541.00)(0.502,2.204){23}{\rule{0.121pt}{1.568pt}}
\multiput(1321.27,541.00)(13.000,51.745){2}{\rule{0.350pt}{0.784pt}}
\multiput(1335.48,596.00)(0.502,2.527){23}{\rule{0.121pt}{1.784pt}}
\multiput(1334.27,596.00)(13.000,59.298){2}{\rule{0.350pt}{0.892pt}}
\multiput(1348.48,659.00)(0.502,2.562){25}{\rule{0.121pt}{1.812pt}}
\multiput(1347.27,659.00)(14.000,65.238){2}{\rule{0.350pt}{0.906pt}}
\put(815.0,418.0){\rule[-0.175pt]{6.263pt}{0.350pt}}
\end{picture}
\end{center}
\caption{\label{fig:instable}
  The $\mu$ dependence of $V_{eff}$
  of the conventional approach in
  the $\alpha$ phase. It has two minima at non-zero $\mu$. Here
  $\Lambda$ is the Euclidean momentum cutoff that defines the model.
  }
\end{figure}
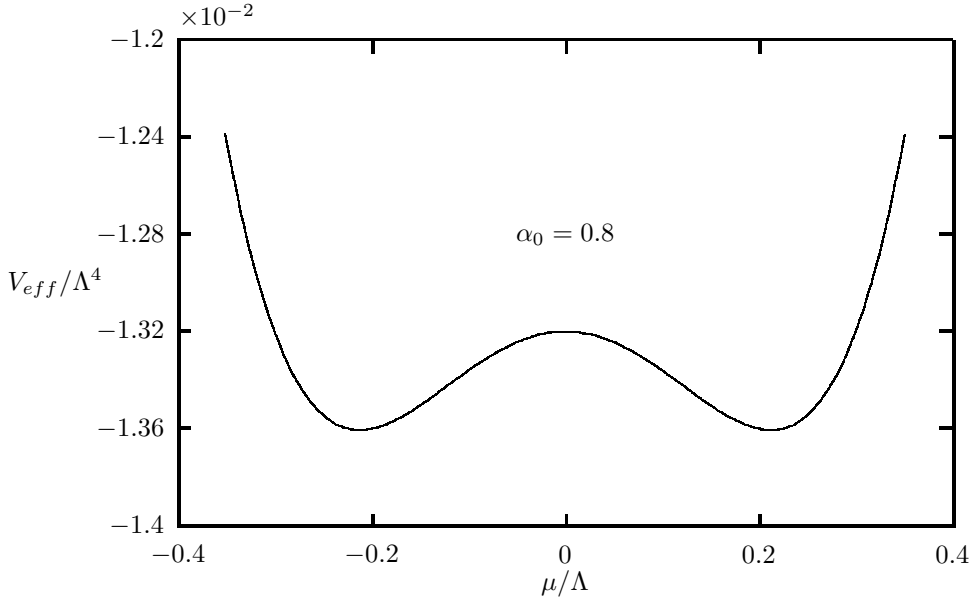

It can be shown that if the covariant integration procedure is
adopted, then $\partial^2 V_{\mbox{\scriptsize eff}}/\partial\mu^2
|_{\mu=0} < 0$ as long as $\sigma\ne 0$ and $D\ge 2$, which implies
that the $\sigma\ne 0$ state with $\overline\rho=0$ is not stable
against fermion number fluctuations. Such a conclusion is independent
of any cutoff scheme (smooth or sharp) and cutoff value for $\Lambda$.
The dependence of the effective potential on the time component of the
statistical gauge field is plotted in Fig. \ref{fig:instable}
Such a result is both
theoretically and physically unacceptable since it would result in
large CP violation in the strong interaction vacuum state which is not
observed. 

\section{The statistical blocking effect}

  It appears that the phase in which $\sigma_{vac}\ne 0$ has a tendency to 
generate finite density of fermion. Such an unexpected behavior may reflect
the fact that some thing is missing in our formalism used so far. 

\begin{figure}[h]
\epsfbox{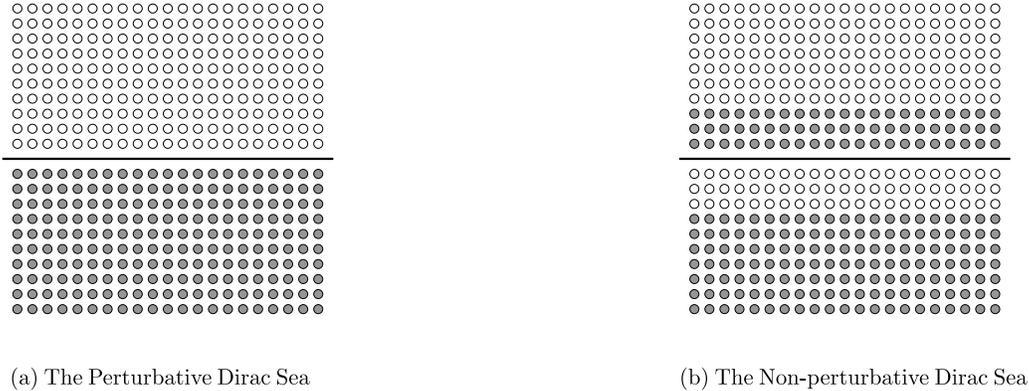}
\caption{\label{fig:upsea}
  The perturbative and non-perturbative Dirac sea for the bare particles. 
The hollow circles represent the empty states.
}
\end{figure}

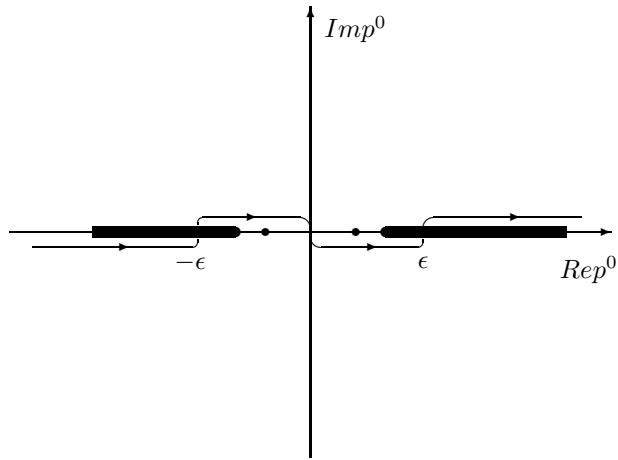
\begin{figure}[h]
\unitlength=1.00mm
\linethickness{0.4pt}
\begin{picture}(120.00,70.00)(0,42)
\put(40.00,80.00){\vector(1,0){80.00}}
\put(80.00,50.00){\vector(0,1){60.00}}
\put(117.00,75.00){\makebox(0,0)[cc]{$Rep^0$}}
\put(86.00,107.00){\makebox(0,0)[cc]{$Imp^0$}}
\put(90.00,79.25){\rule{24.00\unitlength}{1.50\unitlength}}
\put(51.00,79.25){\rule{19.00\unitlength}{1.50\unitlength}}
\put(90.00,80.00){\circle*{1.50}}
\put(70.00,80.00){\circle*{1.50}}
\put(86.00,80.00){\circle*{1.00}}
\put(74.00,80.00){\circle*{1.00}}
\put(43.00,78.00){\line(1,0){20.00}}
\put(97.00,82.00){\line(1,0){19.00}}
\put(67.00,82.00){\line(1,0){11.00}}
\put(82.00,78.00){\line(1,0){11.00}}
\put(63.00,79.00){\oval(4.00,2.00)[rb]}
\put(67.00,81.00){\oval(4.00,2.00)[lt]}
\put(93.00,79.00){\oval(4.00,2.00)[rb]}
\put(97.50,80.50){\oval(5.00,3.00)[lt]}
\put(76.50,80.00){\oval(7.00,4.00)[rt]}
\put(82.00,81.00){\oval(4.00,6.00)[lb]}
\put(53.00,78.00){\vector(1,0){3.00}}
\put(70.00,82.00){\vector(1,0){3.00}}
\put(85.00,78.00){\vector(1,0){4.00}}
\put(102.00,82.00){\vector(1,0){6.00}}
\put(64.00,76.00){\makebox(0,0)[cc]{$-\epsilon$}}
\put(95.00,76.00){\makebox(0,0)[cc]{$\epsilon$}}
\end{picture}
\caption{\label{fig:upcontour}
  The $p^0$ integration contour for the
  effective action for transition amplitudes between states in which both
  the fermion and antifermion states with absolute value of their
  energy below $\varepsilon$
  filled. The thick lines extending to
  positive and negative infinity represent the
  branch cuts of the logarithmic function. 
}
\end{figure}

  An inspection of the Folk space structure of the vacuum state in the
$\sigma_{vac} \ne 0$ phase reveals the problem. It is know that in
such a case, the vacuum state is condensed with {\em finite density}
of fermion--antifermion pairs, namely,
\begin{equation}
     \overline n_{Q\overline Q} = \lim_{V_3\to\infty} {\overline
     N_{Q\overline Q} \over V_3} \ne 0
\label{finit-nqqbar}
\end{equation}
with $N_{Q\overline Q}$ the number of the $Q\overline Q$ pairs and
$V_3$ the spatial volume of the system respectively.  
This is because
\begin{equation}
  \sigma_{vac} \sim \bra{vac}\overline \psi \psi \ket{vac}
\end{equation}
so
\begin{equation}
  \ket{vac} = C_0 \ket{0} + C_1\ket{Q\overline Q} + C_2\ket{Q\overline Q
             Q \overline Q} + \ldots
\end{equation}
with the number of $Q\overline Q$ pairs proportional to $V_3$ if
$\sigma_{vac}\ne 0$.  Like the Fermi sea in the condensed matter
system of fermions, these fermion--antifermion pairs can produce the
statistical blocking effects inside of the system. Such statistical
blocking effects can not be progressively generated by including higher
order loops in a perturbative expansion of the effective action. This
is because the perturbative terms correspond to a modification of the
properties of the system due to finite number of fermions at any
specific order of the perturbation expansion. An effect of
Eq. \ref{finit-nqqbar} can only be accounted for by non-perturbative
means. I proposed a modification of the causal structure of the theory
to include the statistical blocking effect \cite{fd-th1} in a similar
way as the chemical potential do to the theory for the finite density
situations. Fig. \ref{fig:upsea}.b is a graphical representation of
the non-perturbative sea used for our asymptotic ensemble in which the
maximum filled energy level in the positive energy states is controlled
by the statistical block parameter $\varepsilon$.  
It can be shown that if Fig. \ref{fig:upsea}.b is chosen, the energy
integration contour $C_{Phys}$ has to be distorted to ${\cal
C}_{Phys}$ drawn in
Fig. \ref{fig:upcontour}

The effective potential $V_{eff}$ depends, in addition to other
conventional ones, the statistical blocking parameter $\varepsilon$.
Eq. \ref{Veff1} should be replaced by \cite{fd-th1}
\begin{equation}
V_{\mbox{\scriptsize eff}}(\sigma,\mu,\varepsilon) 
= i N_g \int_{{\cal C}_{Phys}} 
{\frac{d^Dp}{(2\pi)^D}}\left [ \ln
\left ( 1-{\frac{\sigma^2 }{p_+^2}} \right ) + \ln \left ( 1-{\frac{\sigma^2 
}{p_-^2}} \right ) \right ] + {\frac{1}{4 G_0}} \sigma^2 + 
{N_g\over 4\pi^2} \left ( \mu^4 + 2\varepsilon^4 + 12 \mu^2 \varepsilon^2 
\right )
\label{Veff3}
\end{equation}
which should be evaluated in the Euclidean space in a way that
preserves the analytic properties of Fig. \ref{fig:upcontour} and
{\em cutted off covariantly}.

\begin{figure}[h]
\epsfbox{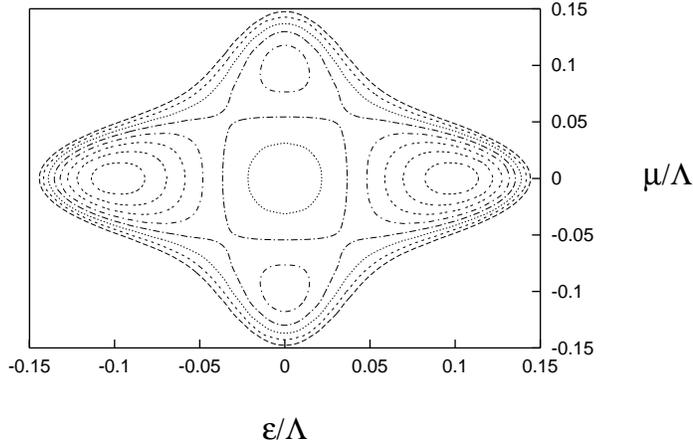}
\caption{\label{fig:veffcntr1} The contour plot of the equal
$V_{\mbox{eff}}$ in the phase in which $D=4$,
$\sigma/\Lambda=1.0$. The point $\mu=0$ and $\varepsilon=0$ is
not stable both in the $\pm \varepsilon$ direction and in the $\pm
\mu$ direction. The absolute minima of the effective potential is
located at about $\mu=0$ and $\varepsilon/\Lambda=\pm 0.1$.  There is
also a pair of local minima located at $\mu/\Lambda = \pm 0.1$ and
$\varepsilon= 0$.}
\end{figure}

The stability of the resulting effective potential for $D=4$ 
is represented in terms of 
equal $V_{eff}$ contour in Fig. \ref{fig:veffcntr1} with a nonvanishing 
$\sigma/\Lambda=1$ and $\Lambda\sim 1$ $GeV$ is the covariant cutoff.
The effective potential is not stable if both $\mu$ and $\varepsilon$ are
zero. The minima at finite $\mu$ are now local minima rather than global.
The global minima are located  at finite $\varepsilon$ and vanish $\mu$ at which
point $\partial^2 V_{\mbox{\scriptsize
eff}}/\partial\mu^2  > 0$ and $\partial^2 V_{\mbox{\scriptsize
eff}}/\partial\varepsilon^2 > 0$ which means that they are stable.  

Therefore, the statistical blocking parameter cure the CP problem mentioned 
above. Some of the implications of having a non-vanishing $\varepsilon$ in
the $\sigma_{vac}\ne 0$ phase of the vacuum is discussed in
Refs. \cite{fd-th1,fd-th2}.

\section{The Properties of Color Superconductivity}

The ground states of the strong interaction could have different
phases from the chiral symmetry breaking phase or the $\alpha$-phase
\cite{lettB,annPhy,Paps-a,Paps-b,Paps-c}.  They are characterized by a
condensation of diquarks causing color superconductivity. Such a
possibility is interesting because it may be realized in the early
universe, in astronomical objects and events, in heavy ion collisions,
inside nucleons \cite{Paps-b,nuc-pap,nuc-pap2} and nuclei, etc.. For the
possible scalar diquark condensation in the vacuum, a half bosonized
model Lagrangian is introduced \cite{Model-I,fd-th1}, which reads
\begin{eqnarray}
{\cal L}_I & = & {\frac{1}{2}} \overline \Psi\left [i{\rlap\slash\partial}
-\sigma- i\vec{\pi}\cdot \vec{\tau}\gamma^5 O_3-\gamma^5 {\cal A}_c\chi^c
O_{(+)}-\gamma^5 {\cal A}^c\overline\chi_c O_{(-)} \right
]\Psi -{\frac{1}{4 G_0}} (\sigma^2 + \vec{\pi}^2) + {\frac{1}{2
G_{3^{\prime}}}} \overline\chi_c \chi^c ,  \label{Model-L-1}
\end{eqnarray}
where $\sigma$, $\vec{\pi}$, $\overline\chi_c$ and $\chi^c$ are auxiliary
fields with $(\chi^c)^{\dagger} = - \overline\chi_c$ and $G_0$, $
G_{3^{\prime}}$ are coupling constants of the model. ${\cal A}_c$ and ${\cal 
A}^c$ $(c=1,2,3)$ act on the color space of the quark; they are ${\cal A}
_{c_1c_2}^c = -\epsilon^{cc_1c_2}$ ${\cal A}_{c,c_1c_2} = \epsilon^{cc_1c_2}$
with $\epsilon^{abc}$ ($a,b = 1,2,3$) the total antisymmetric Levi--Civit\'a
tensor. Here $O_{(\pm)}$ are raising and lowering operators respectively in
the upper and lowering 4 components of $\Psi$.

This model has two non-trivial phases. The vacuum expectation of
$\sigma$ is non-vanishing with vanishing $\chi^2 \equiv-
\overline\chi_c \chi^c$ in the $ \alpha$-phase. The vacuum state in
the $\alpha$-phase is condensed with quark-antiquark pairs. The vacuum
expectation of $\sigma$ is zero with finite $\chi^2$ that
spontaneously breaks the $U(1)$ statistical gauge symmetry in the
second phase, which is called the $\omega$-phase. There is a
condensation of correlated scalar diquarks and antidiquarks in the
color $ \overline 3$ and $3$ states in the $\omega$-phase of the
vacuum state.

\begin{figure}[h]
\epsfbox{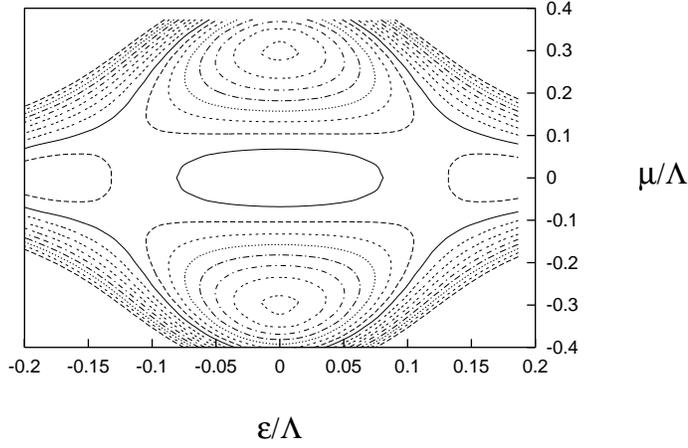} 
\caption{\label{fig:veffcntr21} The contour plot of the equal
$V_{\mbox{eff}}$ in the phase in which 
$\chi_{vac}/\Lambda=1.0$. The point $\mu=0$ and $\varepsilon=0$ is
also not stable both in the $\pm \varepsilon$ direction and in the $\pm
\mu$ direction. The local minima of the effective potential is
located at about $\mu=0$ and $\varepsilon/\Lambda=\pm 0.15$.  There is
also a pair of absolute minima located at $\mu/\Lambda = \pm 0.3$ and
$\varepsilon= 0$.}
\end{figure}

\begin{figure}[h]
\epsfbox{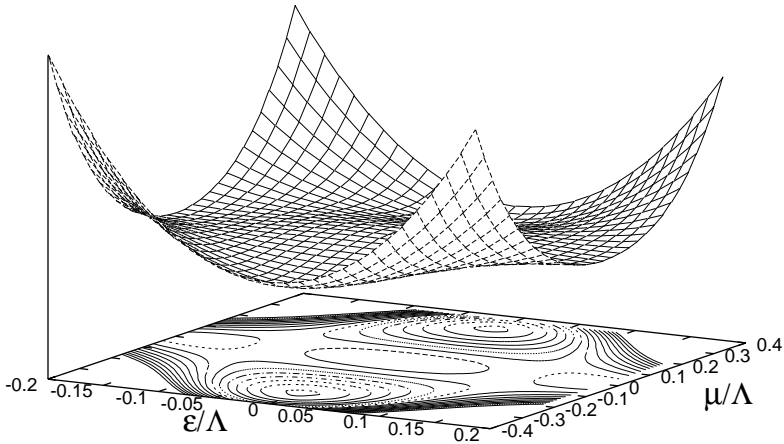} 
\caption{\label{fig:veffcntr22} The 3D plot of 
$V_{\mbox{eff}}$ in the phase in which 
$\chi_{vac}/\Lambda=1.0$.}
\end{figure}

Since diquarks and antidiquarks are condensed in the $\omega$-phase,
it is expected that an exchange of the role of $\varepsilon$ and $\mu$
occurs. It is found to be indeed true: there are also two sets of
minima for the effective potential, the first set is the one with
finite $\mu=\pm\mu_{vac}$ and $ \varepsilon=0$ and the second set
corresponds to $\mu=0$ and finite $ \varepsilon=\pm\varepsilon_{vac}$
in the $\omega$-phase of the model. But here the absolute minima of
the system in the $\omega$-phase correspond to the first set of
solutions, in which the CP invariance and baryon number conservation
are spontaneously violated due to the presence of a finite vacuum
$\mu^\alpha_{vac}$. The stability of the resulting effective
potential is represented in Figs. \ref{fig:veffcntr21} and
\ref{fig:veffcntr22}

\begin{figure}[h]
\epsfbox{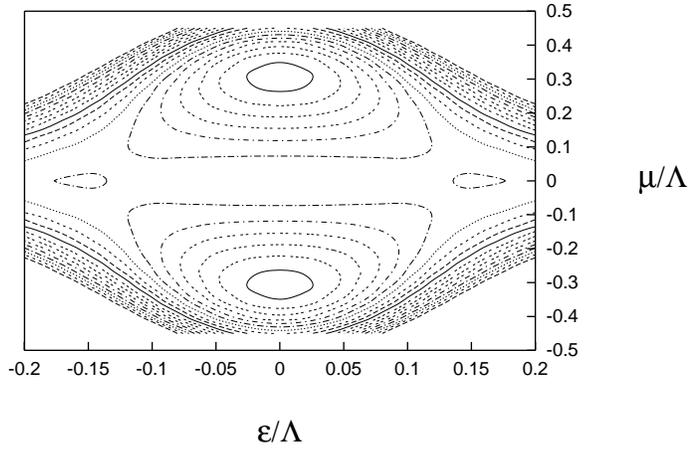} 
\caption{\label{fig:veffcntr31} The contour plot of the equal
$V_{\mbox{eff}}$ in the phase in which 
$\sqrt{\phi^2_{vac}}/\Lambda=1.0$. The point $\mu=0$ and $\varepsilon=0$ is
also not stable both in the $\pm \varepsilon$ direction and in the $\pm
\mu$ direction. The local minima of the effective potential is
located at about $\mu=0$ and $\varepsilon/\Lambda=\pm 0.15$.  There is
also a pair of absolute minima located at $\mu/\Lambda = \pm 0.3$ and
$\varepsilon= 0$.}
\end{figure}

\begin{figure}[h]
\epsfbox{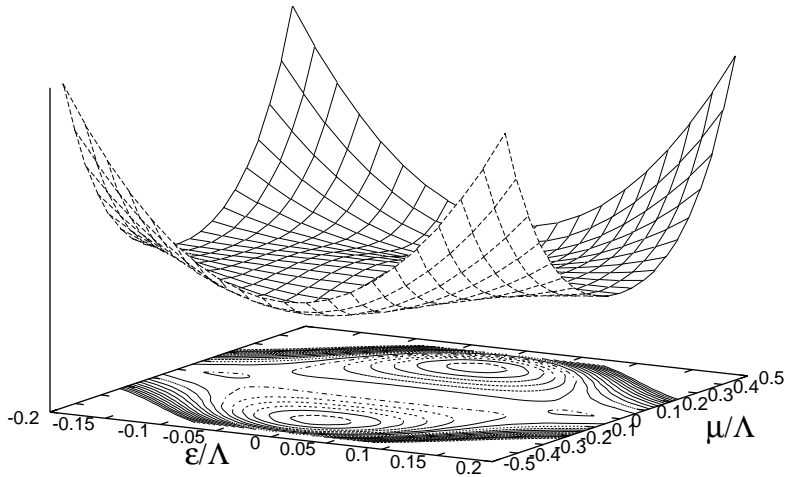} 
\caption{\label{fig:veffcntr32} The 3D plot of 
$V_{\mbox{eff}}$ in the phase in which 
$\sqrt{\phi^2_{vac}}/\Lambda=1.0$. }
\end{figure}

This conclusion is also applicable to the $\beta$-phase of $p$-wave
diquark condensation models
\cite{lettB,annPhy} with vector fermion pair and antifermion pair
condensation, in which the chiral symmetry $SU(2)_L\times SU(2)_R$ is
also spontaneously broken down. It is shown in
Fig. \ref{fig:veffcntr31} and \ref{fig:veffcntr32}

Therefore, the statistical blocking parameter $\varepsilon$ and the
primary statistical gauge field $\mu^\alpha$ provide a more complete
set of macroscopic parameters to describe the condensation of
fermionic particles in a relativistic system. Their physical meaning
is also clear. The statistical gauge field characterizes the
condensation of particle (pairs), which has net fermion number.
The color superconducting phase is caused by condensation of fermion
pairs, which implies that fermion pairing or antifermion pairing are 
favored against the fermion--antifermion pairing. The matter
antimatter tends to separate in the color superconducting phase
resulting in genesis of matter. The statistical blocking parameter
characterizes the concentration of fermion--antifermion pairs with
no net fermion number. The chiral symmetry breaking phase, on
the other hand, favors the fermion--antifermion pairing.
It is not surprising that $\varepsilon\ne 0$ in the chiral symmetry
breaking phase.

\section{Summary}

    In summary, it is important to take into account, among others
ones discussed in Refs. \cite{fd-th1,fd-th2}, of the
statistical blocking effect in a localized finite density theory.

\section*{Acknowledgements}

    I would like to thank Professor Hisakazu Minakata for inviting me
to the ``TMU-Yale Symposium on Dynamics of Gauge Fields''. This work
is supported by NNSF of China under contract 19875009.

\end{document}